\documentclass[reprint,showpacs,pra,nofootinbib,amsmath,amssymb]{revtex4-1}

\usepackage{graphicx}
\usepackage{dcolumn}
\usepackage{bm}
\usepackage{amsmath}
\usepackage{color}
\usepackage{CJK}

\allowdisplaybreaks[2]

\begin{document}
\begin{CJK*}{UTF8}{gbsn}

\title{Controllable linear $\pi$-phase modulation in a thermal atom vapor without diffraction or absorption}

\author{Lida Zhang(\CJKfamily{gbsn}张理达)}

\author{J\"{o}rg Evers}
\affiliation{Max-Planck-Institut f\"{u}r Kernphysik, Saupfercheckweg 1, D-69117 Heidelberg, Germany}

\date{\today}

\begin{abstract}
A scheme is proposed to achieve substantial controllable phase modulation for a probe field propagating through a thermal atomic vapor in double-$\Lambda$ configuration. The phase modulation is based on the linear susceptibility of the probe field, paraxial diffraction is eliminated by exploiting the thermal motion of atoms, and residual absorption is compensated via an incoherent pump field. As a result, a strong controllable uniform phase modulation without paraxial diffraction is achieved essentially independent of the spatial profile or the intensity of the probe field. This phase shift can be controlled via the intensities of the control or the incoherent pump fields. A possible proof-of-principle experiment in alkali atoms is discussed.
\end{abstract}

\maketitle

\end{CJK*}

\section{Introduction}               
Photons are ideal information carriers for information science and telecommunication applications. However, the processing of the encoded information requires interactions, which are more challenging to implement~\cite{kok}. An important example are phase gates, such as controlled $\pi$-phase modulations~\cite{chen2013,Venkataraman2013,Reiserer2014,Tiecke2014}. In order to achieve such phase shifts, there have been extensive studies by utilizing nonlinear effects enhanced by quantum coherences and interferences. For example, self-\cite{wang2001,sheng2011} or cross- \cite{rebic2004,joshi2005,li2008,shiau2011} phase modulation based on Kerr effect have been proposed using electromagnetically induced transparency (EIT)~\cite{schmidt1996,schmidt1998,PhysRevLett.84.1419,niu2005}, spontaneously generated coherences~\cite{niu2006} or active Raman gain~\cite{deng2007,zhu2010} media, not only in gaseous-phase such as atomic alkali atoms, but also in solid-state media including optical fibers~\cite{islam1987,shtaif1998,Venkataraman2013}, quantum wells~\cite{sun2006,jin2013}, and superconducting qubits~\cite{
hu2011,rebic2009,kumar2010}. Several schemes have been experimentally tested in cold~\cite{shiau2011,PhysRevLett.108.173603,chen2013} or thermal atomic systems~\cite{wang2001,kang2003,li2008,li2013}, where  small nonlinear phase shifts up to the order of one radian are achieved. Since nonlinear effects are usually weak, typically long propagation distances are required. Furthermore, these effects sensitively depend on the intensities of the involved laser fields. Thus, the paraxial diffraction which leads to energy attenuation and spatial distortion, has to be taken into account and eventually deteriorates the entire processes. Moreover, diffraction 
also gives rise to an additional nonuniform phase shift over the transverse plane perpendicular to the propagation direction, disturbing the already diffraction-weakened nonlinear phase shift.

\begin{figure}[b]
\centering
\setlength{\dbltextfloatsep}{-10.0mm}
\includegraphics[width=8.cm]{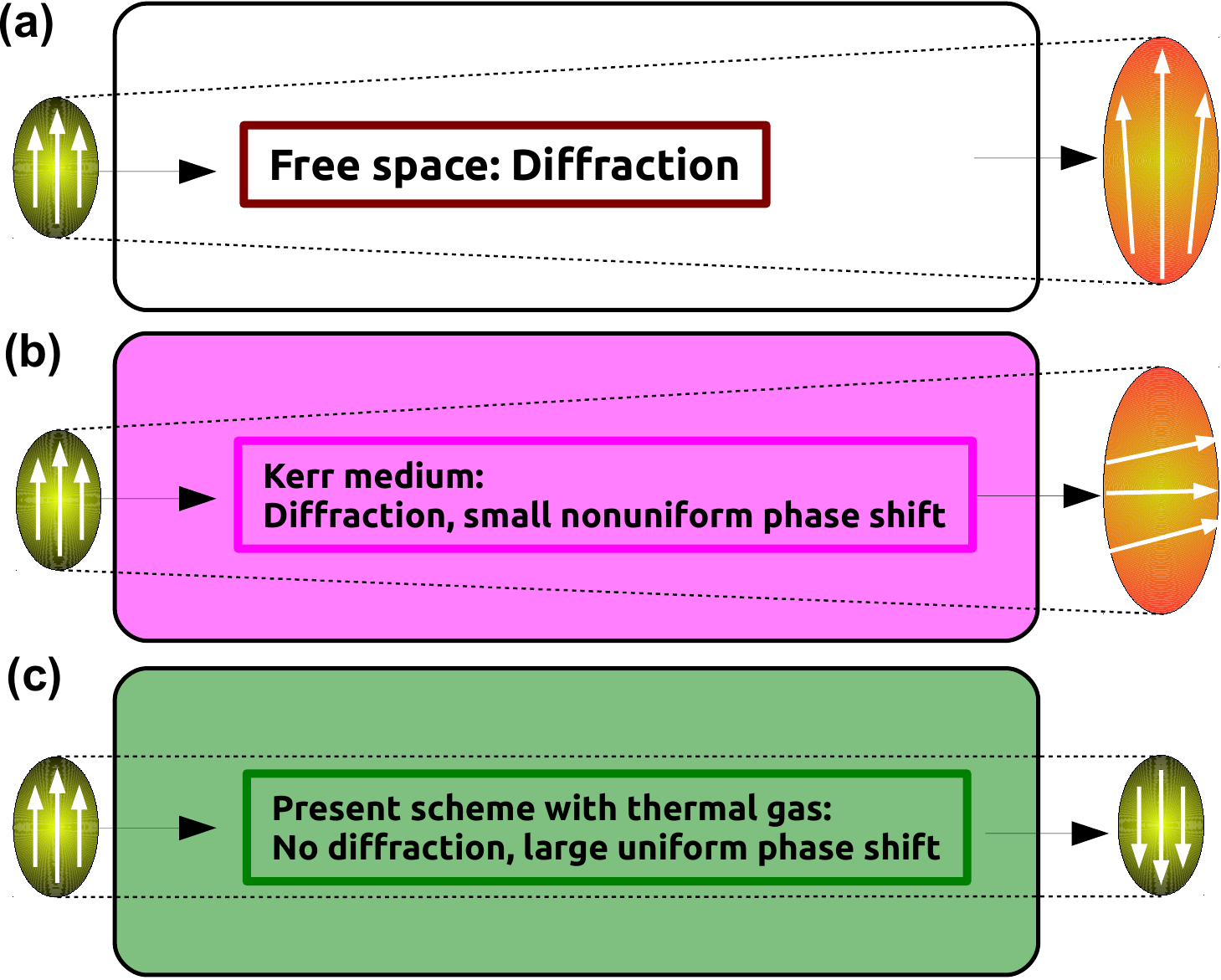}
\caption{(Color online) Schematic illustration for phase modulation acquired by a laser field throughout propagation in different media. The arrows indicate the polarization of the laser field. (a) In free space, diffraction leads to spatial spreading and a small nonuniform phase distribution in the transverse plane. 
(b) In a nonlinear Kerr medium, spatial spreading and nonuniform phase distortion due to diffraction occur. (c) In the setup discussed here, a strong spatially uniform phase shift is achieved, together with cancellation of paraxial diffraction.} 
\label{fig1}
\end{figure}

\begin{figure}[b]
\centering
\setlength{\dbltextfloatsep}{-10.0mm}
\includegraphics[width=8.5cm]{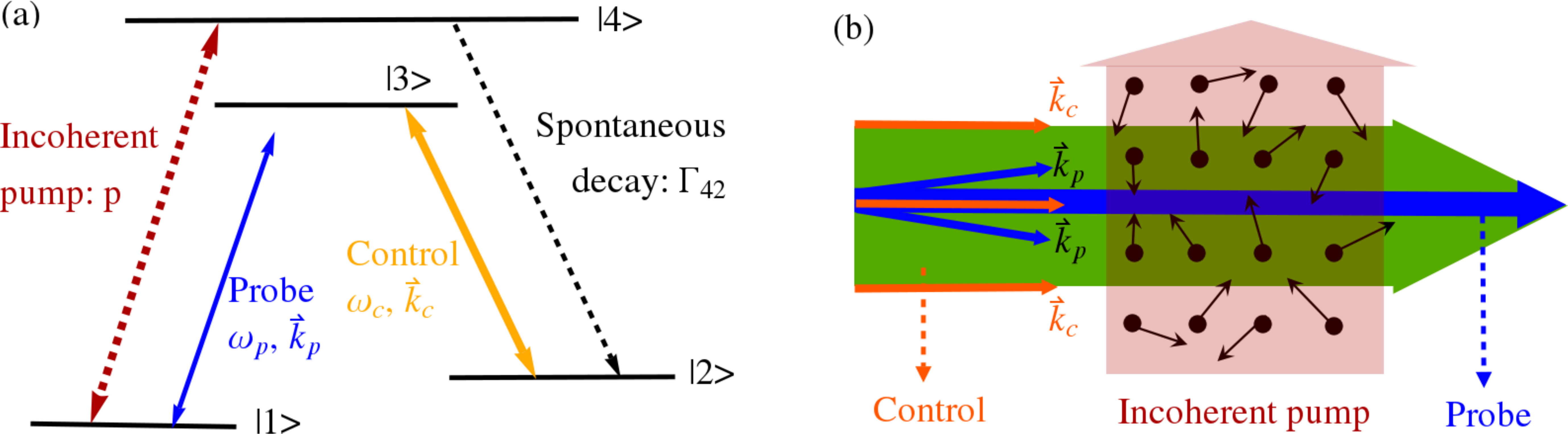}
\caption{(Color online) (a) Atomic level scheme. The four-level double-$\Lambda$ system interacts with co-propagating probe and control fields and a two-way incoherent pump field. The field configuration is sketched in (b). Due to the atomic motion, the paraxial diffraction can be  eliminated.}
\label{fig2}
\end{figure}

Here, we propose a setup which allows to achieve a controllable uniform $\pi$-phase shift over a propagation distance of a fraction of the Rayleigh length. The phase modulation is based on the linear medium susceptibility and therefore does not depend on the intensity of the laser beam. The setup furthermore cancels paraxial diffraction by exploiting the thermal motion of atoms, such that no diffraction-induced nonuniform phase distribution over the transverse plane arises in the propagation. These advantages are illustrated in Fig.~\ref{fig1}. 
The atomic medium configuration is shown in Fig.~\ref{fig2}(a), and consists of a four-level double-$\Lambda$ level scheme interacting with the probe field, a control laser field, and an incoherent two-way pump field. The lower $\Lambda$ subsystem is in electromagnetically induced transparency configuration. Due to atomic motion and collisions, for a negative two-photon detuning between the probe and the control fields, paraxial diffraction for the probe can be exactly canceled, as initially proposed theoretically~\cite{firstenberg2007,firstenberg2008,firstenberg2009a,PhysRevA.89.013817} and later demonstrated experimentally~\cite{firstenberg2009b,firstenberg2013}. In essence, the elimination of diffraction is achieved, since each component of the probe field in the transverse momentum ($\textbf{k}_{\perp}$) space couples stronger with atoms moving in the opposite direction in the transverse plane, and is effectively dragged back towards the main axis. The non-zero two-photon detuning at the same time leads 
to a  linear constant dispersion acting on the probe field, inducing the desired phase modulation throughout the propagation. Furthermore, the single-photon absorption due to the two-photon detuning can be compensated by the gain induced by the incoherent pump, since it will pump out the populations in the ground state and then redistribute the populations among the four states. Via the coherent control field, atomic coherences will be generated between states $|2\rangle$ and $|3\rangle$, which leads to constant gain for the probe field. Altogether, the probe field experiences a controllable phase shift, e.g., of $\pi$, but essentially without any other changes to the beam properties.

This article is organized as follows. In Sec.~II, we first discuss the propagation equation that governs the dynamics of the probe field, and then present the theoretical model. Next we calculate the linear susceptibility of the thermal atomic medium which leads to phase modulation and elimination of diffraction for the probe field. In Sec.~III, we discuss our main results based on numerical simulations. In Sec.~IV, discuss and summarize our results.

\section{Theoretical considerations}

\subsection{Propagation dynamics and origin of the phase shift}  
The propagation dynamics of the probe beam is governed by  Maxwell's equations, which in the paraxial regime and in momentum space can be written as~\cite{firstenberg2009a}
\begin{equation}
\left(\frac{\partial}{\partial z}+i\frac{\text{k}_{\perp}^2}{2\text{k}_{p}}\right)\Omega_{p}(\textbf{k}_{\perp},z) = i\frac{\text{k}_{p}}{2}\chi(\textbf{k}_{\perp}) \Omega_{p}(\textbf{k}_{\perp},z)\,.
\label{propagation}
\end{equation} 
Here, $\Omega_{p}(\textbf{k}_{\perp},z)$ is the Fourier transform of the slowly varying envelope of the probe field  $\Omega_{p}(\textbf{r}_{\perp},z)$ in the transverse plane $(x,y)$ perpendicular to the propagation direction $z$, and $\chi(\textbf{k}_{\perp})$ is the linear susceptibility of the thermal atomic medium  in momentum space. Paraxial diffraction of the probe beam throughout its propagation originates from the second term $\sim \text{k}_{\perp}^2$ on the left hand side. In the following, we will calculate the linear susceptibility $\chi(\textbf{k}_{\perp})$ of the thermal medium in which the dispersion $\text{Re}[\chi]$ can be approximated as $\text{Re}[\chi]=c_{0}+c_{1}\text{k}_{\perp}^2$ around the resonance, while the absorption remains essentially constant and can even  be tuned to zero. Under these conditions, the diffraction can be canceled by the quadratic part $c_{1}\text{k}_{\perp}^2$ of the linear dispersion $\text{Re}[\chi]$~\cite{firstenberg2008,firstenberg2009b,PhysRevA.89.013817}. 
At the same time, the remaining constant part $c_{0}$ of $\text{Re}[\chi]$ gives rise to a phase modulation to the probe field such that $\pi$-phase flips can be  achieved overs short propagation lengths. In other words, the probe field acquires a uniform $\pi$-phase modulation while experiencing neither diffraction nor absorption as it propagates through the thermal medium. Furthermore, since the phase modulation is based on linear effects, it does not require a strong intensity of the probe field which is usually needed for nonlinear phase modulation to obtain large phase shifts.

\subsection{\label{sec-susc}Susceptibility of the thermal medium}
We next describe the scheme to realize the desired linear susceptibility. The four-level double-lambda  scheme is shown in Fig.~\ref{fig2}(a), while the spatial light field configuration is shown in (b).  The transition $|1\rangle \leftrightarrow |3\rangle$ is driven by a probe field with Rabi frequency $\Omega_{p}(\textbf{r},t)$, wavevector $\textbf{k}_{p}$ and detuning $\Delta_{p}=\omega_{p}-\omega_{31}$. The control field with Rabi frequency $\Omega_{c}(\textbf{r},t)$, wavevector $\textbf{k}_{c}$ and detuning $\Delta_{c}=\omega_{c}-\omega_{32}$ drives the transition $|2\rangle \leftrightarrow |3\rangle$.
Additionally, we apply a two-way incoherent pump field $\text{p}(\textbf{r})$ to transition $|1\rangle \leftrightarrow |4\rangle$. As a consequence of this pump field, the atoms initially residing in the ground state $|1\rangle$ will be redistributed among all four atomic states. Together with the control field, atomic coherences between the two states $|2\rangle$ and $|3\rangle$ will be generated already in the absence of the probe field. This  coherences $\rho^{(0)}_{23}$ leads to gain for the probe field. As a result, the overall absorption for the probe field can be controlled to zero or even turned negative, by simply tuning the intensity of the incoherent pump field. The level structure can, for example, be realized in the hyperfine structure of the D1 line of $^{87}\text{Rb}$. In our numerical analysis, we choose the magnetic sublevels 5 $^{2}S_{1/2}, F=1,m_{F}=0$ and $ F=2,m_{F}=2$  as the two lower states $|1\rangle$ and  $|2\rangle$, while 5 $^{2}P_{1/2}, F=2,m_{F}=1$ and $F=1,m_{F}=1$ act as the 
two upper states $|3\rangle$ and  $|4\rangle$, respectively. 

For the analysis, we  follow the approach introduced in \cite{firstenberg2008,PhysRevA.89.013817} to calculate the linear susceptibility for the probe field in the thermal atomic medium under certain approximations: The control field and the incoherent pump field are taken as plane waves, the probe field is assumed to be much weaker than the control fields such that it can be treated as a perturbation to the system, the slowly-varying envelope and paraxial approximations are applied for the probe field, and  the Dicke limit is assumed. The latter approximation is satisfied if  the residual Doppler shift for the two-photon Raman transition $\Delta\text{k}\cdot\text{v}_{\text{th}}$ is much smaller than the combination of the collision rate $\gamma_{c}$ and the incoherent pump rate $\text{p}$, i.e., $\Delta\text{k}\cdot\text{v}_{\text{th}}\ll \text{p}/2+\gamma_{c}$. Those approximations considerably simplify the calculation of the susceptibility, but nevertheless are compatible with state-of-the-art experiments.
 In order to concentrate on the main results, here we only give the final expression for the linear susceptibility, while the detailed procedures to calculate the susceptibility are summarized in the Appendix. We find that the linear susceptibility for the probe field in momentum space can be written as
\begin{align}
\chi(\textbf{k}_{\perp})&=i\alpha\left(\rho^{(0)}_{11}-\rho^{(0)}_{33}+\frac{\Gamma_{c}(\rho^{(0)}_{11}-\rho^{(0)}_{33})+i\Omega_{c}\rho^{(0)}_{23}}{i\Delta-\Gamma_{1}-D\text{k}_{\perp}^2}\right)\,,
\label{susceptibility} 
\end{align}
where $\alpha = 3\lambda_{p}^3\Gamma_{31}K_{31}n_{0}/(8\pi^2)$, with $\lambda_{p}$ being the wavelength of the probe field, $\Gamma_{31}$  the spontaneous decay rate from state $|3\rangle$ to $|1\rangle$, and $n_{0}$ the atomic density. $K_{31}$, which is defined in Eq.~(\ref{def-k31}) in the Appendix, is related to the single-photon spectrum for the probe field. Close to resonance, it can be approximated as a real number~\cite{firstenberg2008}. $\Delta=\Delta_{p}-\Delta_{c}$ is the two-photon detuning of the Raman transition, while $\rho_{ij}^{(0)}$ are the zero-order populations ($i=j$) or coherences ($i\neq j$) for single atoms at rest, which are governed by the incoherent pump and control fields. The other parameters are defined as 
\begin{subequations}
\begin{align}
 D &= \frac{\text{v}_{\text{th}}^{2}}{\gamma_{c}+\frac{\text{p}}{2}+\gamma_{21}-i\Delta}\,, \\
 \Gamma_1 &= \Gamma_{c}+\frac{\text{p}}{2}+\gamma_{21}\,, \\
 \Gamma_c &= K_{31}\Omega_{c}^2\,,
\end{align}
\end{subequations}
where $\gamma_{21}$ is the dephasing rate between the two lower states, and $\Gamma_c$ can be interpreted as power broadening due to the control field.

\subsection{Elimination of diffraction}
From  Eq.~(\ref{susceptibility}) we find that the linear susceptibility depends on the square of $\text{k}_{\perp}$.  In the simple case without incoherent pump field ($p=0$) and in the absence of the probe field, all atoms are in state $|1\rangle$, i.e., $\rho^{(0)}_{11}=1$. Then the linear susceptibility simplifies to 
\begin{align}
  \chi(\textbf{k}_{\perp})&=i\alpha\left(1+\frac{\Gamma_{c}}{i\Delta-\Gamma_{0}-D_{0}\text{k}_{\perp}^2}\right)\,,
\label{simplified}
 \end{align}
with $\Gamma_{0}=\Gamma_{c}+\gamma_{21}$ and $D_{0}=\text{v}_{\text{th}}^{2}/(\gamma_{c}+\gamma_{21}-i\Delta)$. In the regime $\text{k}_{\perp}\ll\text{k}_{0}$ with $\text{k}_{0}=\sqrt{\Gamma_{0}\gamma_{c}/\text{v}_{\text{th}}^2}$, Eq.~(\ref{simplified}) can be expanded in $\text{k}_{\perp}^2$ as
\begin{align}
\label{chi-exp}
  \chi(\textbf{k}_{\perp})&=i\alpha \left(1-\frac{\Gamma_{c}}{2\Gamma_{0}}\right)-\frac{\alpha\Gamma_{c}}{2\Gamma_{0}}+\frac{\alpha\Gamma_{c}}{2\Gamma_{0}}\frac{\text{k}_{\perp}^2}{\text{k}^2_{0}}+O(\text{k}_{\perp}^4)\,,
\end{align}
where we have chosen $\Delta=-\Gamma_{0}$ to cancel the $\text{k}_{\perp}^2$ dependence of the absorption, which may lead to diffusion for the probe field. Plugging Eq.~(\ref{chi-exp}) into the propagation equation Eq.~(\ref{propagation}), we find that in the region $\text{k}_{\perp}\ll\text{k}_{0}$ the paraxial diffraction is eliminated by the linear dispersion under the condition 
\begin{align}
\frac{1}{\text{k}^2_{p}}=\frac{\alpha\Gamma_{c}}{2\Gamma_{0}\text{k}^2_{0}}\,.
\end{align} 
Assuming this condition, the propagation equation for the probe field becomes
\begin{align}
\frac{\partial}{\partial z}\Omega_{p}(\textbf{k}_{\perp},z) = -\frac{\text{k}_{p}\alpha}{2} \left[\left(1-\frac{\Gamma_{c}}{2\Gamma_{0}}\right) + i\frac{\Gamma_{c}}{2\Gamma_{0}}\right]\Omega_{p}(\textbf{k}_{\perp},z)\,.
\label{simplifiedpropagation}
\end{align}
The right hand side of Eq.~(\ref{simplifiedpropagation}) contains a constant term $i\Gamma_c/(2\Gamma_0)$ responsible for the phase modulation throughout the propagation. We will show later that a $\pi$-phase flip can be obtained throughout a propagation of a few Rayleigh lengths. However, we further find in the right hand side of Eq.~(\ref{simplifiedpropagation}) that there is strong absorption for the probe field, which originates from the non-zero two-photon detuning chosen to obtain the quadratic dependence of the susceptibility on $\textbf{k}_{\perp}$. Therefore, the probe field will be severely attenuated, making the phase flip useless.

\subsection{Elimination of absorption}
We found that this single-photon absorption can be  compensated by  atomic coherences $\rho_{23}$ which lead to gain for the probe field. The simplest way to induce the atomic coherences is to apply a two-way incoherent pump field, as shown in Fig.\ref{fig2}(a). In the presence of this pump field and for a resonant control field ($\Delta_{c}=0$), the steady-state populations and coherences in zeroth order of $\Omega_{p}$ evaluate to
\begin{subequations}
\label{coherences} 
\begin{align}
\rho_{11}^{(0)}&= 4 \Gamma_{31} (\text{p}+\Gamma_{4}) \Omega_{c}^2/N\,, \\
\rho_{33}^{(0)}&= 4 \text{p}\Gamma_{42} \Omega_{c}^2/N\,, \\
\rho_{23}^{(0)}&= -i2 \text{p}\Gamma_{3}\Gamma_{42} \Omega_{c} / N\,,
\end{align}
\end{subequations}
with $N=\text{p} \Gamma_{3}^2 \Gamma_{42}+4 [2 \text{p} (\Gamma_{31}+\Gamma_{42})+\Gamma_{31} \Gamma_{4}] \Omega_{c}^2$.
We find from Eqs.~(\ref{coherences}) that the population in the ground state $\rho_{11}^{(0)}$ gradually decreases as the incoherent pump rate $\text{p}$ increases, whereas $\rho^{(0)}_{33}$ and $\rho^{(0)}_{23}$ increase. This means that the strong single-photon absorption proportional to the population difference $\rho^{(0)}_{11}-\rho^{(0)}_{33}$ can be overcome for a suitable choice of $\text{p}$. Moreover, it can further be  compensated by the induced atomic coherences $\rho^{(0)}_{23}$. To see this in more detail, we expand Eq.~(\ref{susceptibility}) in $\text{k}_{\perp}$ and obtain 
\begin{align}
\chi(\textbf{k}_{\perp})=c_{0}+c_{1}\frac{\text{k}^{2}_{\bot}}{\text{k}^{2}_{1}}+O(\text{k}^{4}_{\bot})\,,
\label{expand}
\end{align}
where 
\begin{subequations}
\begin{align}
c_{0}&=i\alpha\left(\rho^{(0)}_{11}-\rho^{(0)}_{33}+\frac{\Gamma_{c}(\rho^{(0)}_{11}-\rho^{(0)}_{33})+i\Omega_{c}\rho^{(0)}_{23}}{i\Delta-\Gamma_1}\right)\,, \\
c_{1}&=i\alpha\Gamma_1\gamma_{c1}\frac{\Gamma_{c}(\rho^{(0)}_{11}-\rho^{(0)}_{33})+i\Omega_{c}\rho^{(0)}_{23}}{(\gamma_{c1}-i\Delta)(i\Delta-\Gamma_1)^2}\,, \\
\gamma_{c1}&=\gamma_{c}+\frac{\text{p}}{2}+\gamma_{21} \,,\\
\text{k}_{1}&=\sqrt{\frac{\Gamma_{1}\gamma_{c1}}{\text{v}^{2}_{\text{th}}}} \nonumber \\
&=\frac{1}{\text{v}_{\text{th}}}\sqrt{(\gamma_{c}+\frac{\text{p}}{2}+\gamma_{21})(\Gamma_{c}+\frac{\text{p}}{2}+\gamma_{21})}\,. 
\end{align}
\label{constants}
\end{subequations}
In the regime $\text{k}_{\bot}\ll\text{k}_{1}$, the constant term $c_0$ leads to uniform absorption and dispersion, while the $\text{k}_{\perp}^2$-dependent term proportional to $c_1$ results in atomic-motion induced absorption and elimination of diffraction. As already found for the case without pump field, a suitable choice
for the two-photon detuning 
\begin{align}
\Delta=-\sqrt{\frac{\gamma_{c1}}{\gamma_{c1}+2\Gamma_{1}}}\Gamma_{1} 
\end{align}
can be employed to remove the $\text{k}_{\perp}^2$-dependent absorption which gives rise to diffusion. 
Furthermore, diffraction is canceled if
\begin{align}
&\frac{1}{\text{k}^{2}_{p}}=\frac{\text{Re}[c_{1}]}{\text{k}^{2}_{1}} =-\alpha\Gamma_{1}\gamma_{c1}\Delta  \nonumber \\
&\times \frac{[\Gamma_{c}(\rho^{(0)}_{11}-\rho^{(0)}_{33})+i\Omega_{c}\rho^{(0)}_{23}](\Gamma_{1}^{2}-\Delta^{2}+2\gamma_{c1}\Gamma_{1})}{(\gamma^{2}_{c1}+\Delta^{2})(\Gamma_{1}^2+\Delta^{2})^2\text{k}^{2}_{1}}\,.
\label{zerodiffraction}
\end{align}
Combined with Eqs.~(\ref{propagation}) and (\ref{susceptibility}), we finally obtain the propagation equation
\begin{align}
\frac{\partial}{\partial z}\Omega_{p}(\textbf{k}_{\perp},z) = i\frac{\text{k}_{p}\alpha}{2}c_{0}\Omega_{p}(\textbf{k}_{\perp},z)\,.
\label{simplifiedpropagation1}
\end{align}
This equation is readily solved by
\begin{equation}
\Omega_{p}(r_{\perp},z) =\Omega_{p}(r_{\perp},z=0)\,e^{\frac{i}{2} \text{k}_{p}\alpha c_{0}z} \,.
\label{solution}
\end{equation}
Eq.~(\ref{solution}) shows that the probe field will propagate through the thermal medium preserving its spatial profile, i.e., without diffraction. By changing the incoherent pump and the control field, the constant absorption proportional to $\text{Im}[c_{0}]$ can be tuned to zero or even negative values. As a result, at zero absorption, as desired the probe field only experiences a uniform phase shift proportional to $\text{Re}[c_{0}]$ during the propagation while all other beam properties remain the same.

\section{Analysis of the phase modulation}

\subsection{Linear phase shift without absorption or diffraction}

From Eq.~(\ref{solution}) we found that uniform phase shifts can be achieved without affecting any other property of the probe field, such as the spatial profile or the power for appropriate parameter choices. 
Then this phase shift can be  written as 
\begin{align}
\phi_{s}= \frac{1}{2}\text{k}_{p}\alpha\text{Re}[c_{0}]z\,,
\label{phase}
\end{align}
which does not depend on the shape or the intensity of the probe field, and thus is universal.
The derivation of Eq.~(\ref{solution}) relied on several approximations. In order to investigate the validity of these approximations, we next simulate the propagation dynamics of the probe field in the thermal medium by numerically solving Eq.~(\ref{propagation}) together with Eq.~(\ref{susceptibility}). The result is subsequently Fourier transformed back to real space. For a proof-of-principle demonstration, the incident probe field is assumed to have a Gaussian spatial profile
\begin{align}
 \Omega_{p}(x,y,z=0)=\Omega_{p0}e^{-(x^2+y^2)/(2w_{p}^2)}\,.
\end{align}
The initial width of the Gaussian probe is set to $w_{p}=100\mu m$, which is within the paraxial regime. We assume parameters of the D1 line of $^{87}\text{Rb}$ for the thermal atoms, since it is the most common atomic species with which related experiments have been conducted up to now. We further choose a propagation distance of one Rayleigh length $z_{R}=2\pi w_{p}^{2}/\lambda_{p}$ which for the present parameters evaluates to about $7.90$cm. 

\begin{figure}[tbp]
\centering
\setlength{\dbltextfloatsep}{-10.0mm}
\includegraphics[width=8cm]{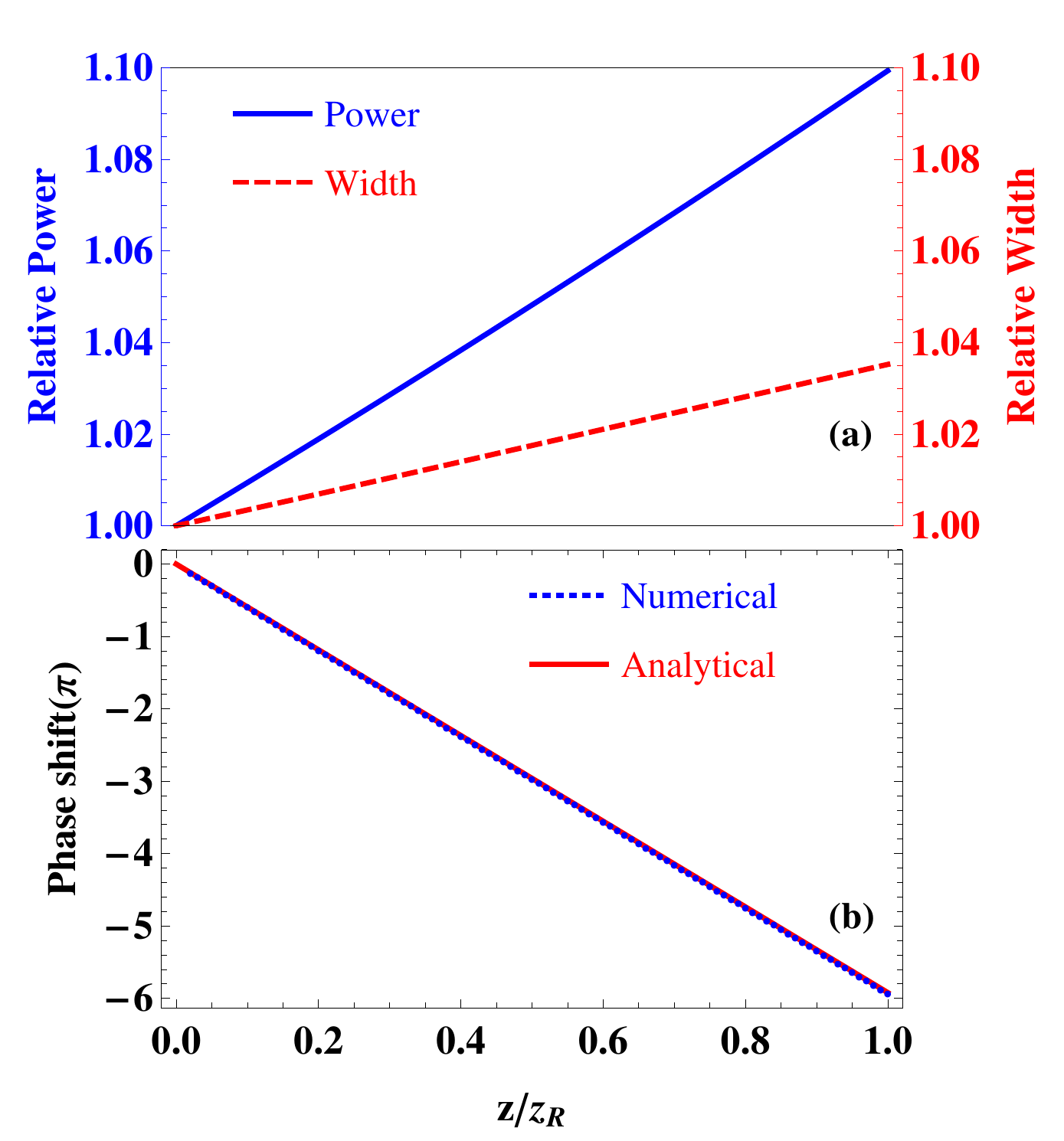}
\caption{(Color online) (a) Power and spatial width of the Gaussian probe beam as function of the propagation distance. Note the small plot range of only few percent relative change, suggesting that the probe field power and width remain approximately unchanged. (b) The accumulated phase shift as a function of propagation distance. $\pi$-phase flips can be achieved already in a small fraction of the Rayleigh length $z_R$.  Parameters are: $n_{0}=1.5\times 10^{12}\text{cm}^{-3}, \lambda_{p}=795\text{nm}, T=300K,\text{v}_{\text{th}}=240 \text{m/s},\Delta\text{k}=22.8\text{m}^{-1},\gamma_{c}=2000\Delta\text{k}\cdot\text{v}_{\text{th}},\Gamma=2\pi\times 5.75 \text{MHz}, \Gamma_{31}=\Gamma/4,\Gamma_{32}=\Gamma/6,\Gamma_{41}=\Gamma/12,\Gamma_{42}=\Gamma/2,\gamma_{21}=0.001\Gamma_{31},\Omega_{c}=1.4\Gamma_{31}, \text{p}=0.65\Gamma_{31}$.} 
\label{fig3}
\end{figure}

Results are depicted in Fig.~\ref{fig3}. As can be seen from  panel (a), the Gaussian shape of the probe is maintained throughout the propagation in the thermal vapor, except for a small broadening by up to $3.5\%$ due to the residual higher-order diffraction $\sim O(\text{k}^{4}_{\perp})$. The probe is also weakly amplified as the propagation distance increases, since we have chosen a relatively strong incoherent pump field such that the gain from the atomic coherences exceeds the reduced single-photon absorption. Note that more precise tuning of the parameters results in a further reduction of the broadening and amplification. After having established that shape and intensity remain essentially unchanged, we turn to the phase, which is shown in Fig.~\ref{fig3}(b). While propagating through the thermal atoms, a phase shift starting from $0$ up to almost $6\pi$ is imprinted onto the probe beam within the single Rayleigh length. Already after $z_{f}\simeq0.168z_{R}\simeq1.33\text{cm}$, which is  a small 
fraction of the Rayleigh length, a $\pi$-phase flip can be achieved. This phase modulation can readily be understood, since the non-zero two-photon detuning results in a deviation from the EIT resonance, such that a nonzero linear dispersion appears for the probe field. It is important to note that the $\pi$-phase flip is accomplished due to linear effects and therefore independent of the spatial shape and power of the incident probe field. 

\begin{figure}[t!]
\centering
\setlength{\dbltextfloatsep}{-10.0mm}
\includegraphics[width=8cm]{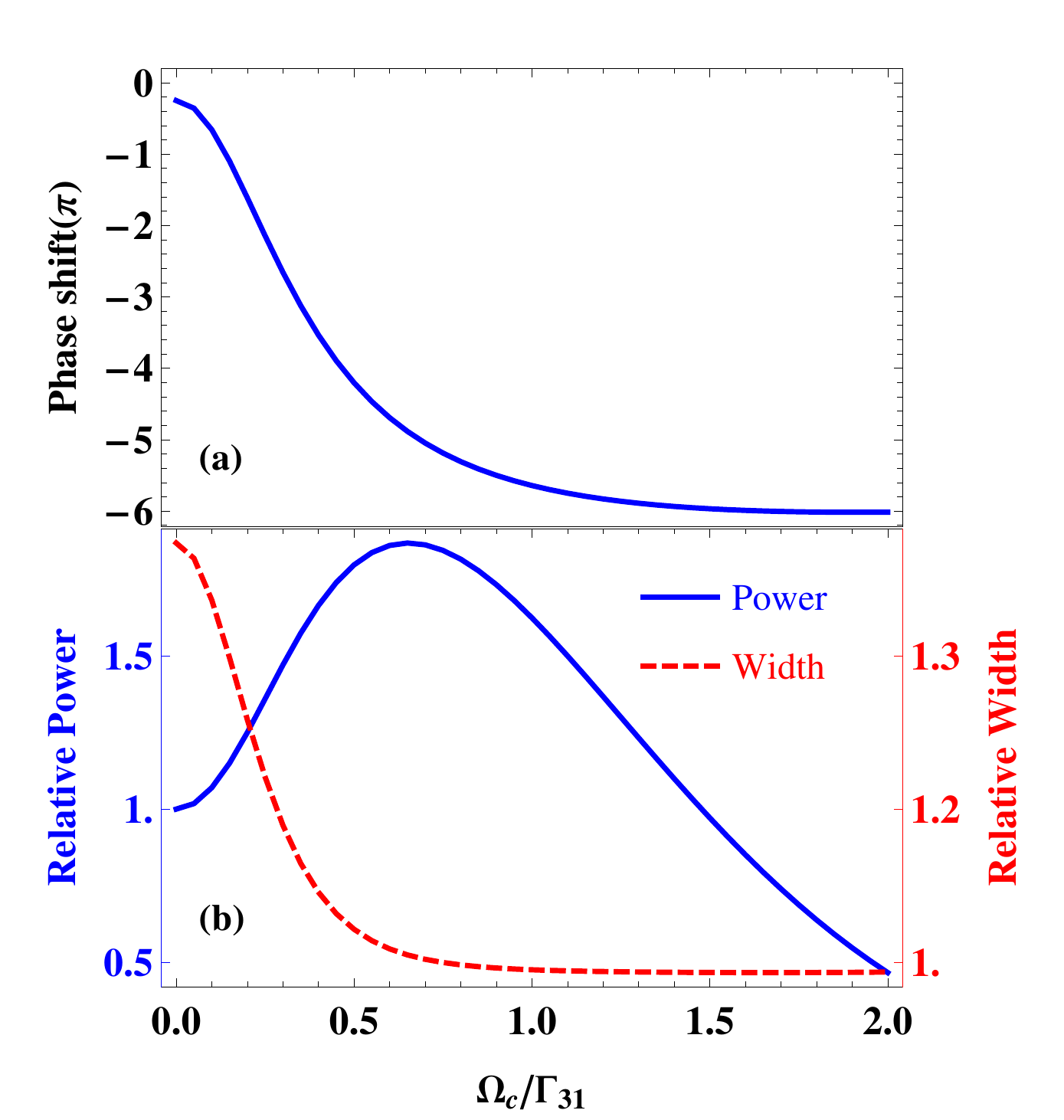}
\caption{(Color online) (a) Phase shift of the transmitted Gaussian probe field against the intensity of the control after propagating one Rayleigh length in the thermal atomic gas. (b) The power and width of the outgoing probe field as a function of the control.  Other parameters are as in Fig.~\ref{fig3}.} 
\label{fig4}
\end{figure}

\begin{figure}[t]
\centering
\setlength{\dbltextfloatsep}{-10.0mm}
\includegraphics[width=8cm]{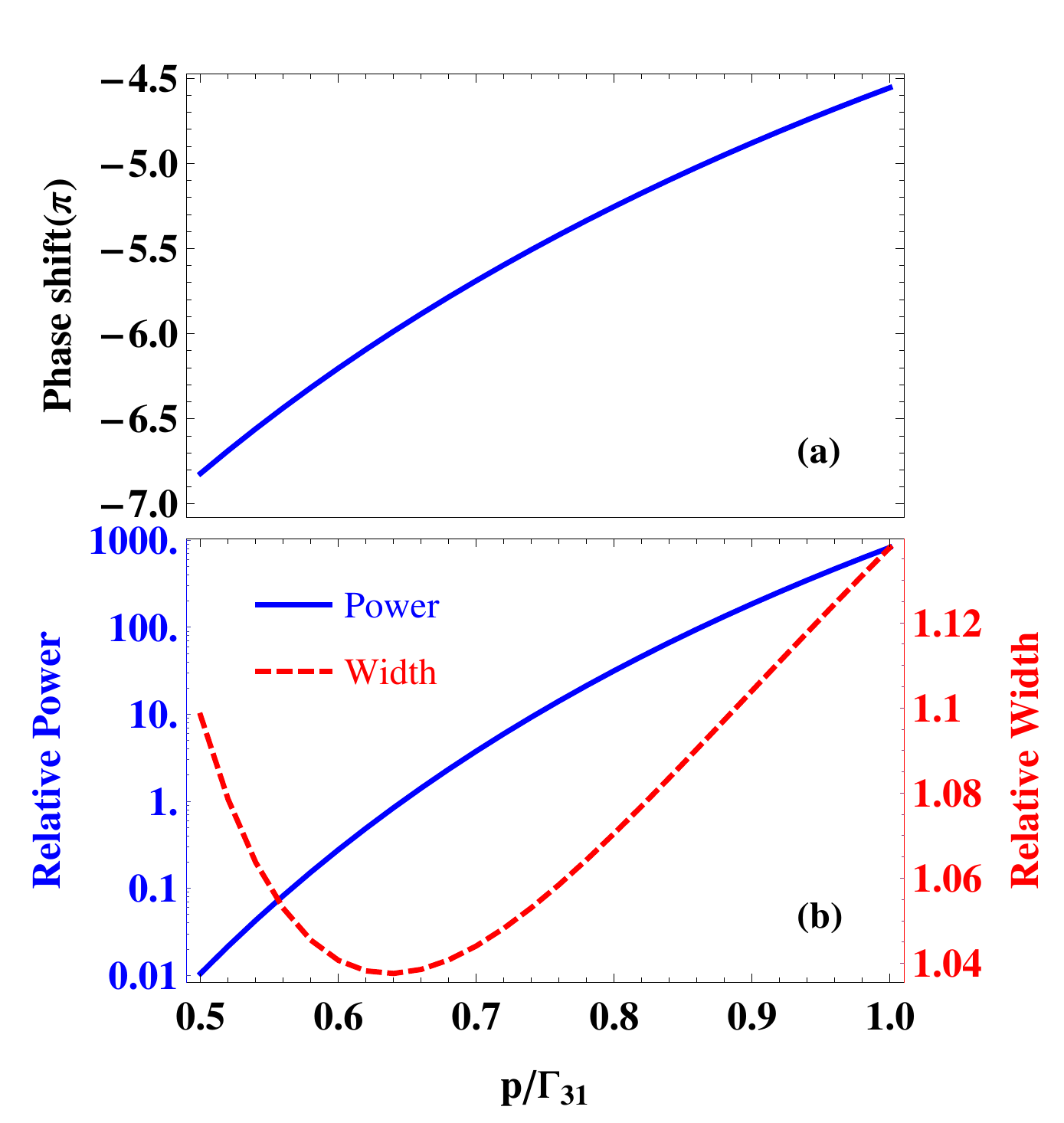}
\caption{(Color online) (a) Phase shift of the output Gaussian probe field as a function of the incoherent pump  rate after propagating one Rayleigh length in the thermal atomic gas. (b) The power and width of the outgoing probe field as function of the incoherent pump rate.  Other parameters are as in Fig.~\ref{fig3}.} 
\label{fig5}
\end{figure}

\subsection{Effect of the control field intensity}
The phase modulation can be controlled not only by the propagation distance, but also by other parameters such as the intensities of the control and pump fields,  as illustrated by Eqs.~(\ref{phase}),~(\ref{constants}a) and~(\ref{coherences}). In this section, we study the effect of the control field intensity. Results are shown in Fig.~(\ref{fig4}) for a Gaussian probe field, which propagates in the thermal medium for a distance equal to one Rayleigh length. In the absence of a coupling field ($\Omega_{c}=0$), since the atoms have been optically pumped out of the ground state $|1\rangle$ to $|2\rangle$ by the incoherent pump, there is no interaction between the atoms and the probe field any more. Then the probe propagation proceeds as in free space. As shown in Fig.~\ref{fig4}(a), the probe field acquires a small phase shift due to the diffraction term, while the power remains the same and the width is broadened to $w(z=z_{R})=\sqrt{2}w_{p}$ as depicted in Fig.~\ref{fig4}(b) by the red dashed line. 
Increasing $\Omega_{c}$, the phase shift can gradually be tuned, due to the $\Omega_{c}$-dependent constant dispersion $\text{Re}[c_{0}]$. At $\Omega_{c}$ exceeding about $1.5\,\Gamma_{31}$, the phase shift becomes approximately independent of $\Omega_{c}$ since $\text{Re}[c_{0}]$ has saturated to its minimum [but maximum absolute value, see Fig.~\ref{fig4}(a)]. 
By dynamically changing the control field parameters, the phase shift can also be switched between two values. For example, a controllable phase shift of $\pi$ could be realized by toggling $\Omega_{c}$ between $0.47\Gamma_{31}$ and $0.7\Gamma_{31}$. The relative phase shift imprinted onto the probe beam between these two intensities is $\pi$. Due to the switching, output probe field intensity slightly changes (relative power increases from 1.76 to 1.86), and the probe beam width is broadened by less  than $8\%$.

\subsection{\label{sec-inc}Effect of the incoherent pump field}
Next to the control field power, the phase shift can also be tuned via the intensity of the incoherent pump field, which we studied next. Results are shown in Fig.~(\ref{fig5}) for a Gaussian probe field propagating in the thermal medium for a distance equal to one Rayleigh length. We find that the absolute value of the acquired phase shift is approximately inversely proportional to the incoherent pump rate $\text{p}$ as shown in Fig.~\ref{fig5}(a). Thus, a $\pi$-phase flip can also be achieved by choosing a suitable pump strength. At lower pump rates, the width of the outgoing probe field decreases with increasing $\text{p}$, and reaches a minimum roughly at $\text{p}=0.65\Gamma_{31}$ at which the condition for diffraction cancellation is satisfied. For other values of $\text{p}$, the paraxial diffraction is either over- or under- compensated, resulting in a broadened width for the output probe. Since the populations and coherences sensitively depend on the incoherent pump power as indicated in Eq.~(\ref{coherences}), the output power of the probe field strongly depends on the incoherent pump rate,  as shown in Fig.~\ref{fig5}(b). 

\subsection{\label{sec-both}Combining control field and incoherent pump field control}
\begin{figure}[t]
\centering
\setlength{\dbltextfloatsep}{-10.0mm}
\includegraphics[width=6.cm]{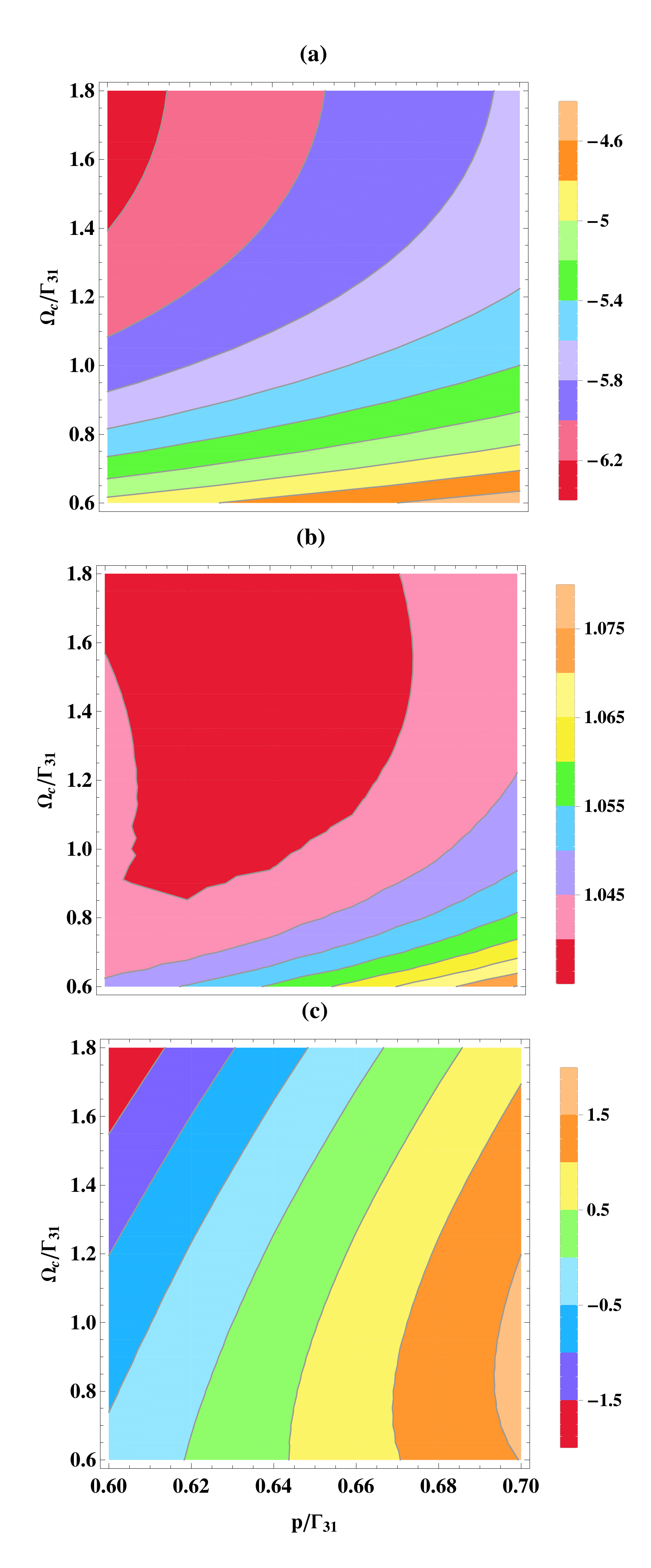}
\caption{(Color online) (a) Phase shift (in unit of $\pi$) of the output Gaussian probe field plotted against the intensities of the control and incoherent pump fields after propagating one Rayleigh length in the thermal atomic gas. A $\pi$-phase flip can be achieved in various ways by tuning the control and incoherent pump simultaneously. (b) and (c) show the corresponding relative power ($\text{ln}[P_{\text{out}}/P_{\text{in}}]$) and width ($w_{\text{out}}/w_{p}$) of the outgoing probe field. Other parameters are the same as in Fig.~\ref{fig3}. } 
\label{fig6}
\end{figure}

More generally, the phase shift can be tuned by simultaneously changing the control field and incoherent pump field intensities. This way, results can further be improved compared to the control of only one of the two variables. An example is shown in Fig.~\ref{fig6} for a Gaussian probe field propagating one Rayleigh length in the thermal medium. Again it can be seen that a controllable $\pi$-phase flip can be easily realized. For example, if an operation with switching between phase shifts of $-6\pi$ and $-5\pi$ is chosen, then the width of the outgoing probe beam remains almost the same for the two values, with residual broadening less than $7\%$. Note that the output power of the probe field is much more sensitive to the incoherent pump rate than to the control field power, since the population redistribution and coherences which lead to reduction of the single-photon absorption crucially depend on the intensity of the incoherent pump.

\begin{figure}[t]
\centering
\setlength{\dbltextfloatsep}{-10.0mm}
\includegraphics[width=8.5cm]{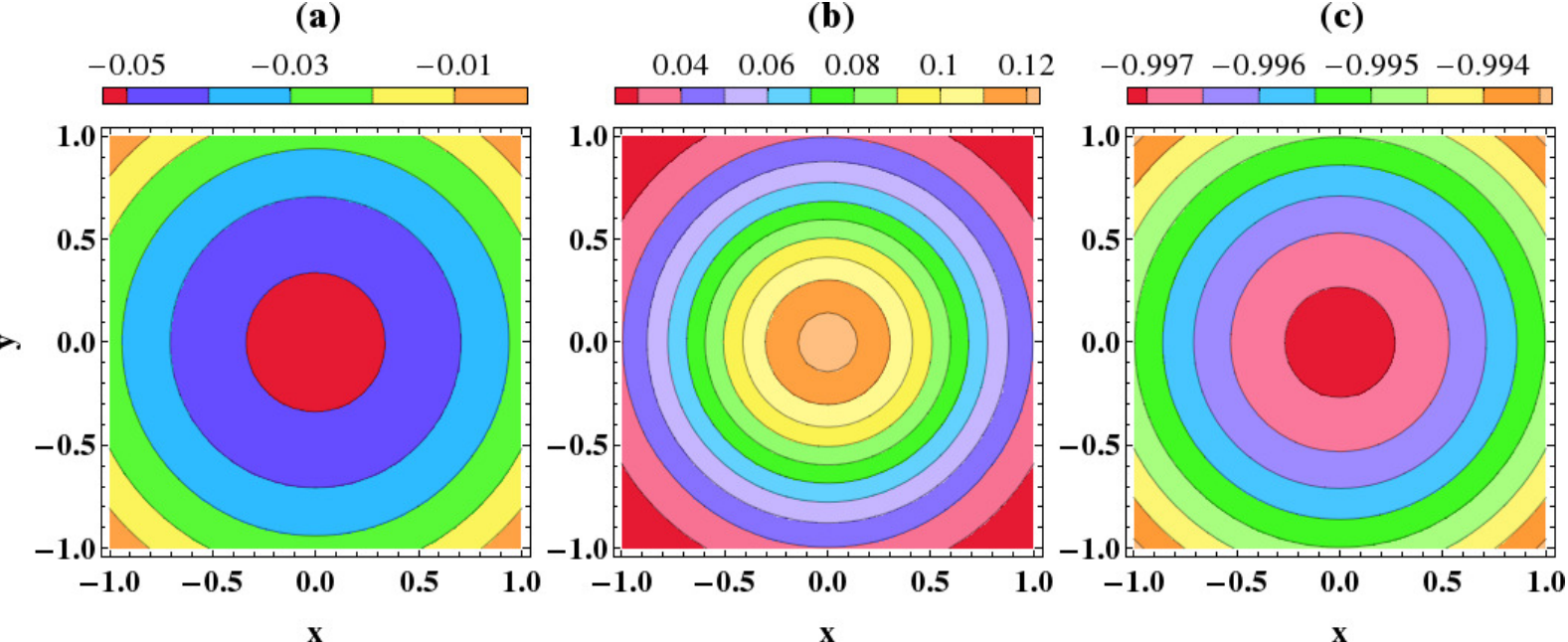}
\includegraphics[width=8.5cm]{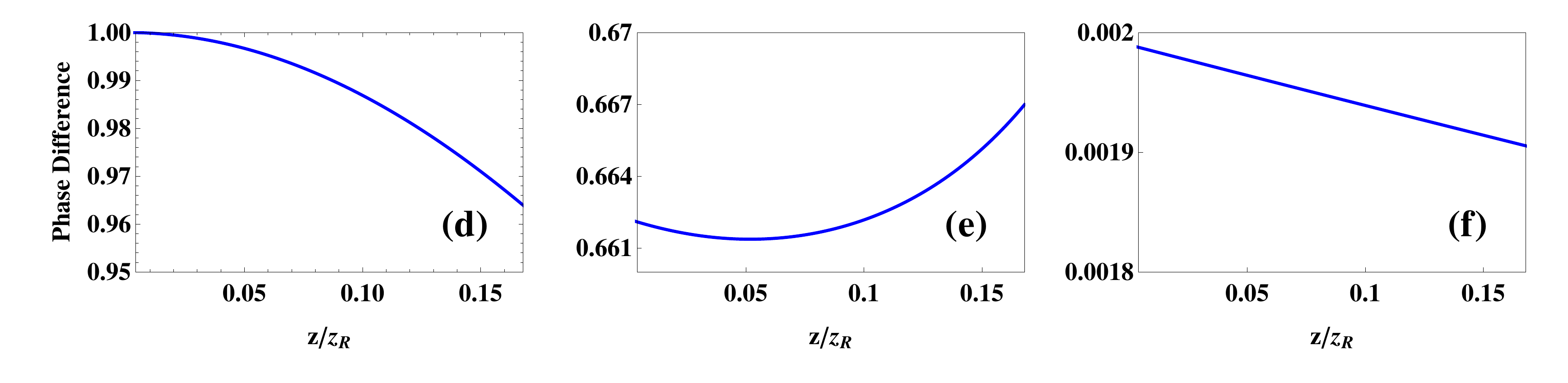}
\caption{(Color online) Spatial phase variation of the probe field in the transverse plane in units of $\pi$ after a propagation distance $z_{f}=0.168z_{R}$. Results are shown for free space (a), a Kerr medium (b), and the thermal vapor considered here (c). The parameters chosen for the Kerr medium are the same as in Fig.(\ref{fig3}) except for $\Omega_{p0}=0.1\Omega_{c}$, $\text{p}=0$, $T = 0K.$ In the thermal vapor, the probe acquires an almost uniform phase $\pi$, as compared to the much smaller nonuniform phase gained in Kerr medium or in free space. (d)-(f) show the phase difference, which is defined as $(\phi[x=y=0]-\phi[x=y=w_{p}])/(\phi[x=y=0]+\phi[x=y=w_{p}])$, as a function of propagation distance for the three cases. The starting points for $z$ in (d)-(f) are $z=0.002z_{R}$, since at $z=0$ the phase difference for all cases should be 0. Other parameters are the same as in Fig.~\ref{fig2}. } 
\label{fig7}
\end{figure}

\section{Spatial uniformity of the phase modulation}
Throughout the calculations up to now, the phase value is extracted at the point of peak intensity of the probe field (i.e., at $x=y=0$). Ideally, the phase imprinted onto the probe field should be uniformly distributed in the plane transverse to the propagation direction, since there is only an overall constant dispersion term affecting the propagation dynamics independent of the probe field intensity. However, the dispersion introduced by the residual diffraction in the region beyond $\text{k}_{\perp}\ll\text{k}_{1}$ and higher-order diffraction $\sim O(\text{k}^{4}_{\perp})$ can lead to small variations in the phase distribution over the transverse plane. This can be further improved by increasing $\text{k}_{1}$ defined in Eq.~\ref{constants}(d). As an example, we calculated the phase distribution over the transverse plane after a propagation distance $z_{f}=0.168z_{R}$ in free space, in a  Kerr medium,  and in our thermal gas. For the Kerr medium, we calculated the nonlinear atomic coefficient $\chi_{3}$ in the same atomic system at $T=0K$ and $p=0$, and simulated the propagation dynamics with $\partial_{z}\Omega_{p}(\textbf{r}_{\bot},z)=i\partial^2_{\textbf{r}_{\bot}}\Omega_{p}(\textbf{r}_{\bot},z) + \chi_{3}|\Omega_{p}(\textbf{r}_{\bot},z)|^2\Omega_{p}(\textbf{r}_{\bot},z)$. Results are shown in Fig.\ref{fig7}(a)-(c). After this propagation distance $z_{f}$, the probe beam has accumulated a nearly uniform phase shift of $\pi$ in the thermal medium, with the phase differences in the transverse plane smaller than $0.2\%$ as shown in Fig.\ref{fig7}(f). In the Kerr medium, the acquired phase modulation is up to $0.13\pi$, and accompanied with a considerable phase difference larger than $66.0\%$ as shown in Fig.\ref{fig7}(e).

\section{Discussions and Conclusions}
In our numerical calculations, we have assumed parameters of the D1 line of $^{87}\text{Rb}$, which was used in previous experiments. But in principle, our scheme can be realized in any thermal atomic system in which a double-$\Lambda$ level structure can be found. However, since we have not considered direct interactions between the atoms~\cite{fleischhaker2010,PhysRevA.71.033802} 
in our scheme,   the atomic gas should be dilute enough to neglect them. In order to allow our scheme to work at lower atomic densities, we found that it is favorable if a large ratio $\Gamma_{42}/\Gamma_{41}$ between spontaneous decay rates  minimizes the population in the upper state $|4\rangle$ which does not interact with the probe field. Moreover, the coupling between the atoms and the probe field should be as strong as possible to further reduce the required atomic density, such that a large dipole moment $\vec\mu_{31}$ is favorable.

As discussed above, we initially employed the two-way incoherent pump field to redistribute the populations and induce atomic coherences already at zeroth order in the probe field, which would lead to reduction of one-photon absorption. Surprisingly, we found that the pump field has several other positive effects beyond our initial purpose. First, the incoherent pump further alleviates the demand for strong collision rates to achieve the Dicke limit. Second, since $\text{k}_{1}$ defined in Eq.~(\ref{constants}d) which sets the transverse wave number scale in Eq.~(\ref{expand}) grows rapidly as the incoherent pump increases, the series expansion in the transverse momentum becomes more accurate with the pump field, resulting in a reduction of the differences in the phase distribution of the probe field across the transverse plane. Third, the pump field introduces one more degree of freedom to control the phase shift of the probe field as discussed in Sec.~\ref{sec-inc} and \ref{sec-both}.

In summary, we have studied the propagation of a probe field through a thermal atomic medium, and have shown that it is possible to imprint large phase shifts onto the probe field together with diffraction cancellation, such that the spatial beam width and the beam intensity remain approximately unchanged. In particular, we have discussed the possibility to imprint a controllable phase flip of $\pi$ onto the field. 
Our scheme is applicable for probe fields with arbitrary spatial profiles within a certain transverse momentum bandwidth, and is independent of the probe field intensity. The phase shift can be controlled via the intensity of the control and incoherent pump fields. For a proof-of-principle demonstration, we discussed a possible experimental implementation using the hyperfine structure of the D1 line in atomic $^{87}\text{Rb}$. In principle, our scheme can be extended to the low-photon level, as long as the noise induced by the gain mechanism remains low enough~\cite{kong2013,kocsis2013}. 

We are grateful for funding by the German Science Foundation (DFG, Sachbeihilfe EV 157/2-1). 

\appendix

\numberwithin{equation}{section}

\section{\label{sec:appendix}Derivation of the linear susceptibility}
In general, the calculation of the linear susceptibility is similar to the procedures introduced in~\cite{firstenberg2008,PhysRevA.89.013817}. Following the theoretical description developed in~\cite{firstenberg2008,PhysRevA.89.013817}, we define a generalized density-matrix distribution function in space and velocity as
\begin{equation} 
\rho(\textbf{r},\textbf{v},t)=\sum\limits_{i} \rho^{i}(t)\:\delta(\textbf{r}-\textbf{r}_{i}(t))\: \delta(\textbf{v}-\textbf{v}_{i}(t))\,.
\label{eq2}
\end{equation}
Here, $\rho^{i}(t)$ is the density matrix for the $i$-th atom. $\rho(\textbf{r},\textbf{v},t)$ can be understood as the probability density to find an atom with internal density matrix $\rho(t)$ at position $\textbf{r}$ and with velocity $\textbf{v}$. Then, the equation of motion of the system can be written as
\begin{align}
\frac{\partial \rho(\textbf{r},\textbf{v},t)}{\partial t} &=- \frac{i}{\hbar} [H, \rho(\textbf{r},\textbf{v},t)] -L\rho(\textbf{r},\textbf{v},t)\nonumber \\
&\quad - \textbf{v}\cdot\frac{\partial \rho(\textbf{r},\textbf{v},t)}{\partial \textbf{r}}  \nonumber \\ 
&\quad -\gamma_{c}\big[\rho(\textbf{r},\textbf{v},t) - R(\textbf{r},t)F(\textbf{v})\big]\,,
\label{eq3}
\end{align}
where $H$ is the Hamiltonian of the system, $L\rho$ represents the relaxation terms including spontaneous decay, dephasing and the incoherent pump, and $\gamma_{c}$ is the collision rate.
In the right hand side of Eq.~~(\ref{eq3}), the first two terms describe the internal quantum-mechanical evolution, while the next two terms characterize the external classical motion including thermal motion and collisions~\cite{firstenberg2008}.
The last contribution contains the density of atoms in internal state $\rho(t)$ per unit volume at position $\textbf{r}$  
\begin{align}
R(\textbf{r},t) = \int\rho(\textbf{r},\textbf{v},t)d\textbf{v}\,, 
\label{eq4}
\end{align}
as well as the Boltzmann distribution of the atom velocities $F(\textbf{v})=\text{Exp}[{-\text{v}^2/\text{v}_{\text{th}}^2}]/\sqrt{\pi}\text{v}_{\text{th}}$
with $\text{v}_{\text{th}}=\sqrt{2k_{b}T/m}$ being the most probable thermal velocity.

The relevant equations of motion for the coherences follow as
\begin{subequations}
\label{eq5}
\begin{align}
&\big(\frac{\partial}{\partial t} + \textbf{v}\cdot\frac{\partial}{\partial \textbf{r}}-i\Delta_{p}+i\textbf{k}_{p}\cdot\textbf{v}+\frac{\text{p}(\textbf{r})}{2}+\frac{\Gamma_{3}}{2}+\gamma_{c}\big)\rho_{31} \nonumber \\
&= i\Omega_{p}(\textbf{r},t)(\rho_{11}-\rho_{33})+i\Omega_{c}(\textbf{r},t) \rho_{21}+\gamma_{c}R_{31}(\textbf{r},t)F(\textbf{v})\,,       \\
&\big(\frac{\partial}{\partial t} + \textbf{v}\cdot\frac{\partial}{\partial \textbf{r}}-i\Delta+i\Delta\textbf{k}\cdot\textbf{v}+\frac{\text{p}(\textbf{r})}{2}+\gamma_{21}+\gamma_{c}\big)\rho_{21} \nonumber \\
&= i\Omega_{c}^{*}(\textbf{r},t) \rho_{31}-i\Omega_{p}(\textbf{r},t)\rho_{23} +\gamma_{c}R_{21}(\textbf{r},t)F(\textbf{v})\,, 
\end{align}
\end{subequations}
where we have abbreviated $\rho_{ij}(\textbf{r},\textbf{v},t)$ as $\rho_{ij}$, and introduced the two-photon detuning $\Delta = \Delta_{p}-\Delta_{c}$ and the wavevector difference $\Delta\textbf{k}=\textbf{k}_{p}-\textbf{k}_{c}$. Here, $\Delta_{i}$ and $\textbf{k}_{i}$ is the detuning and wave vector of the field with Rabi frequency $\Omega_{i}(\textbf{r},t)(i\in p,c)$, $\text{p}(\textbf{r})$ is the incoherent pump rate. We denote the spontaneous emission rate on transition $|i\rangle \to |k\rangle$ as $\Gamma_{ik}$, and the total decay rate of state $|i\rangle$ as  $\Gamma_{i}=\sum_{k}\Gamma_{ik}$. The dephasing between the two ground state is $\gamma_{21}$.

For simplicity, we treat the control field $\Omega_{c}(\textbf{r},t)$ and the pump field $\text{p}(\textbf{r})$ as plane-wave fields, which means $\Omega_{c}(\textbf{r},t)=\Omega_{c}$ and $\text{p}(\textbf{r})=\text{p}$. Then, the steady-state density-matrix distribution function in the zeroth-order of the probe field can be obtained as\cite{firstenberg2008}
\begin{align}
\rho^{(0)}_{ij}(\textbf{r},\textbf{v}) =n_{0}\rho^{(0)}_{ij}F(\textbf{v})\,.
\label{eq6}
\end{align}
Here, $n_{0}$ is the atomic density and $\rho^{(0)}_{ij}$ is the zero-order density matrix element of an atom at rest. 

Under those approximations, the equations of motion for the first-order coherences can be derived from Eq.~~(\ref{eq5})
\begin{subequations}
\begin{align}
&\big(\frac{\partial}{\partial t} + \textbf{v}\cdot\frac{\partial}{\partial \textbf{r}}-i\Delta_{p}+i\textbf{k}_{p}\cdot\textbf{v}+\frac{\text{p}}{2}+\frac{\Gamma_{3}}{2}+\gamma_{c}\big)\rho^{(1)}_{31} \nonumber \\
&= i\Omega_{p}(\textbf{r},t)(\rho^{(0)}_{11}-\rho^{(0)}_{33})n_{0}F(\textbf{v})+i\Omega_{c}(\textbf{r},t) \rho^{(1)}_{21} \nonumber \\
&\quad+\gamma_{c}R^{(1)}_{31}(\textbf{r},t)F(\textbf{v})\,,       \\[2mm]
&\big(\frac{\partial}{\partial t} + \textbf{v}\cdot\frac{\partial}{\partial \textbf{r}}-i\Delta+i\Delta\textbf{k}\cdot\textbf{v}+\frac{\text{p}}{2}+\gamma_{21}+\gamma_{c}\big)\rho^{(1)}_{21} \nonumber \\
&= i\Omega_{c}^{*}(\textbf{r},t) \rho^{(1)}_{31}-i\Omega_{p}(\textbf{r},t)\rho^{(0)}_{23}n_{0}F(\textbf{v}) +\gamma_{c}R^{(1)}_{21}(\textbf{r},t)F(\textbf{v})\,, 
\end{align}
\label{eq8}
\end{subequations}
In order to get a analytical expression for $R^{(1)}_{31}(\textbf{k},\omega)$, which determine the thermal atomic effect on the probe propagation dynamics, we first integrate Eqs.~~(\ref{eq8}b) over velocity
\begin{align}
&\big(\frac{\partial}{\partial t} -i\Delta+\frac{\text{p}}{2}+\gamma_{21}\big)R^{(1)}_{21}(\textbf{r},t) +\big(\frac{\partial}{\partial \textbf{r}}+i\Delta\textbf{k}\big)\textbf{J}_{21}(\textbf{r},t)\nonumber \\
&= i\Omega_{c}^{*}R^{(1)}_{31}(\textbf{r},t)-in_{0}\Omega_{p}(\textbf{r},t)\rho^{(0)}_{23},
\label{eq9}
\end{align}
here we have defined the current density of the density-matrix distribution function
\begin{equation}
\textbf{J}_{ij}(\textbf{r},t)=\int\textbf{v}\rho^{(1)}_{ij}(\textbf{r},\textbf{v},t)d\textbf{v},    \\
\label{eq10}
\end{equation}
in Eq.~~(\ref{eq8}b), when the Dicke limit is satisfied, i.e., $\gamma=\text{p}/2+\gamma_{c}+\gamma_{21}-i\Delta$ is dominant, we can approximately rewrite $\rho(\textbf{r},\textbf{v},t)$ in $\gamma$ to first order
\begin{align}
 \rho^{(1)}_{21}(\textbf{r},\textbf{v},t)&=
\rho^{(1,0)}_{21}(\textbf{r},\textbf{v},t)+\frac{1}{\gamma}\rho^{(1,1)}_{21}(\textbf{r},\textbf{v},t) \nonumber \\
&=R^{(1)}_{21}(\textbf{r},t)F(\textbf{v})+\frac{1}{\gamma}\rho^{(1,1)}_{21}(\textbf{r},\textbf{v},t).
\label{eq11}
\end{align}
We can then find that 
\begin{subequations}
\begin{align}
0&=\int\textbf{v}\rho^{(1,0)}_{21}(\textbf{r},\textbf{v},t)d\textbf{v}, \\
\textbf{J}_{21}(\textbf{r},t)&=\frac{1}{\gamma}\int\textbf{v}\rho^{(1,1)}_{21}(\textbf{r},\textbf{v},t)d\textbf{v}, 
\end{align}
\label{eq12}
\end{subequations}
In Eq.(\ref{eq8}b), we expand $\rho^{(1)}_{21}(\textbf{r},\textbf{v},t)$ as in Eq.~~(\ref{eq11}) and multiply Eq.~~(\ref{eq8}b) by $\textbf{v}$, then integrate over velocity. Using the relations in Eq.~~(\ref{eq12}) and taking the leading term in $\gamma$, the following equation for $\textbf{J}_{21}(\textbf{r},t)$ yields
\begin{align}
\textbf{J}_{21}(\textbf{r},t)&=-D\big(\frac{\partial}{\partial\textbf{r}}+i\Delta\textbf{k}\big)R^{(1)}_{21}(\textbf{r},t)+i\frac{\Omega_{c1}^{*}}{\gamma}\textbf{J}_{31}(\textbf{r},t),
\label{eq13} 
\end{align}
where $D$ is defined as $D=\text{v}_{\text{th}}^{2}/\gamma$. To derive Eq.~~(\ref{eq13}) we have used the relation~\cite{firstenberg2008}
\begin{align}
\int\textbf{v}^2\frac{\partial}{\partial\textbf{r}}R^{(1)}_{21}(\textbf{r},t)F(\textbf{v})d\textbf{v}=\text{v}_{\text{th}}^2\frac{\partial}{\partial\textbf{r}}R^{(1)}_{21}(\textbf{r},t),
\label{eq14} 
\end{align}
substituting $\textbf{J}_{21}(\textbf{r},t)$ in Eq.~~(\ref{eq9}) by Eq.~~(\ref{eq13}) we have
\begin{align}
&\big[\frac{\partial}{\partial t} -i\Delta+\frac{\text{p}}{2}+\gamma_{21}-D\big(\frac{\partial}{\partial\textbf{r}}+i\Delta\textbf{k}\big)^2\big]R^{(1)}_{21}(\textbf{r},t)\nonumber \\
&= i\Omega_{c}^{*}R^{(1)}_{31}(\textbf{r},t)-in_{0}\Omega_{p}(\textbf{r},t)\rho^{(0)}_{23}\nonumber \\
&\quad-i\big(\frac{\partial}{\partial\textbf{r}}+i\Delta\textbf{k}\big)\cdot\frac{\Omega_{c1}^{*}}{\gamma}\textbf{J}_{31}(\textbf{r},t),
\label{eq15} 
\end{align}
the last term containing $\textbf{J}_{31}(\textbf{r},t)$ in Eq.~~(\ref{eq15}) usually can be neglected when $|\Omega_{c}|\ll|\gamma|$. Furthermore, even when this condition is not satisfied, these terms can be still neglected when both the spatial variations $\partial/\partial\textbf{r}$ and $\Delta\textbf{k}$ remains in the transverse directions, perpendicular to $\textbf{k}_{p}$, since $\textbf{J}_{31}(\textbf{r},t)$ is parallel to $\textbf{k}_{p}$~\cite{firstenberg2008}. Then Eq.~(\ref{eq15}) is simplified as
\begin{align}
&\big[\frac{\partial}{\partial t} -i\Delta+\frac{\text{p}}{2}+\gamma_{21}-D\big(\frac{\partial}{\partial\textbf{r}}+i\Delta\textbf{k}\big)^2\big]R^{(1)}_{21}(\textbf{r},t)\nonumber \\
&= i\Omega_{c}^{*}R^{(1)}_{31}(\textbf{r},t)-in_{0}\Omega_{p}(\textbf{r},t)\rho^{(0)}_{23}.
\label{eq16} 
\end{align}
Here we consider the common case of slowly-varying envelope approximation(SVEA), where the temporal and spatial variations in the envelope of probe field are much smaller than the decoherence rate and the wave number. Correspondingly, it results in the SVEA for the spatial-temporal evolution of density-matrix distribution function $R^{(1)}_{31}(\textbf{r},t)$, which leads to
\begin{align}
&\bigg|\frac{\partial}{\partial t}+\textbf{v}\cdot\frac{\partial}{\partial\textbf{r}}\bigg|\ll\bigg|\frac{\text{p}}{2}+\frac{\Gamma_{3}}{2}-i\Delta_{p}+i\textbf{k}_{p}\cdot\textbf{v}\bigg|,
\label{eq17}  
\end{align} 
In Eq.~~(\ref{eq8}a), we can then neglect the temporal and spatial variations, and only take the dominant part of $\rho^{(1)}_{21}(\textbf{r},\textbf{v},t)=R^{(1)}_{21}(\textbf{r},t)F(\textbf{v})$, then integrate over velocities, the expression for $R^{(1)}_{31}(\textbf{r},\textbf{v},t)$ is found
\begin{align}
R^{(1)}_{31}(\textbf{r},t)&=iK_{31}\big[n_{0}(\rho^{(0)}_{11}-\rho^{(0)}_{33})\Omega_{p}(\textbf{r},t)+\Omega_{c}R^{(1)}_{21}(\textbf{r},t)\big],
\label{eq18}
\end{align}
where $K_{31}$ is defined as follows
\begin{subequations}
\label{def-k31}
\begin{align}
G_{31}&=\int\frac{F(\textbf{v})}{\Delta_{p}-\textbf{k}_{p}\cdot\textbf{v}+i(\frac{\text{p}}{2}+\frac{\Gamma_{3}}{2}+\gamma_{c})}d\textbf{v},\\[1mm]
K_{31}&=\frac{iG_{31}}{1-i\gamma_{c}G_{31}},
\end{align}
\label{eq19}
\end{subequations}
when $\Delta_{p}\ll \text{p}/2+\Gamma_{3}+\gamma_{c}$, i.e., near the one-photon resonance where most experiments were done, the imaginary parts of $K_{31}$ are much smaller than its rear part. In the following, we treat $K_{31}$ as real number. 
In the paraxial approximation, the changes in the envelopes along z direction are much smaller than that in the transverse plane, so we may replace $\textbf{r}\rightarrow(\textbf{r}_{\perp},z)$ and $\partial/\partial\textbf{r}\rightarrow\partial/\partial\textbf{r}_{\perp}$ in Eq.~~(\ref{eq16}), and then Fourier transform them from $(\textbf{r}_{\perp},t)$ to $(\textbf{k}_{\perp},\omega)$. We can then immediately find the final expression for $R^{(1)}_{31}(\textbf{k}_{\perp},\omega)$
\begin{align}
R^{(1)}_{31}(\textbf{k}_{\perp},z,\omega)&=iK_{31}n_{0}\Omega_{p}(\textbf{k}_{\perp},z,\omega)
\big(\rho^{(0)}_{11}-\rho^{(0)}_{33} \nonumber \\
&+\frac{\Gamma_{c}(\rho^{(0)}_{11}-\rho^{(0)}_{33})+i\Omega_{c}\rho^{(0)}_{23}}{i(\omega+\Delta)-\Gamma_{1}-D(\textbf{k}_{\perp}+\Delta\textbf{k})^2}\big),
\label{eq20} 
\end{align}
where we have set the power broadening $\Gamma_{c}=K_{31}\Omega^{2}_{c}$ and $\Gamma_{1}=\text{p}/2+\gamma_{21}+\Gamma_{c}$. For a continuous wave, we can set $\omega=0$ in Eq.~~(\ref{eq20}). In the case for a small wavevector difference between $\Omega_{p}$ and $\Omega_{c}$, $\Delta\textbf{k}$ could be neglected, i.e., $\Delta\textbf{k}=0$. 

Finally, we note that the propagation equations for the probe field in momentum space can be written as 
\begin{align}
\left(\frac{\partial}{\partial z}+i\frac{\text{k}_{\bot}^{2}}{2\text{k}_{p}}\right)\Omega_{p}(\textbf{k}_{\bot},z)
&=i\frac{3\lambda_{p}^2\Gamma_{31}}{8\pi}R_{31}(\textbf{k}_{\bot},z)
\label{prop-m}
\end{align}
By comparing Eqs.~(\ref{prop-m}) with Eqs.~(\ref{propagation}), we can then find the expression for the linear susceptibility
\begin{align}
\chi(\textbf{k}_{\perp}) &=
\frac{3\lambda^{3}_{p}\Gamma_{31}}{8\pi^{2}}\frac{R^{(1)}_{31}(\textbf{k}_{\perp},z,\omega=0)}{\Omega_{p} (\textbf{k}_{\perp},z,\omega=0)} \nonumber \\
&=i\alpha\big(\rho^{(0)}_{11}-\rho^{(0)}_{33}+\frac{\Gamma_{c}(\rho^{(0)}_{11}-\rho^{(0)}_{33})+i\Omega_{c}\rho^{(0)}_{23}}{i\Delta-\Gamma_{1}-D\text{k}_{\perp}^2}\big).
\label{eq21} 
\end{align}
where we have $\alpha = 3\lambda_{p}^3\Gamma_{31}K_{31}n_{0}/(8\pi^2)$.

\bibliographystyle{apsrev4-1}
%

\end{document}